\documentclass{aa}
\usepackage{txfonts}
\usepackage{graphicx}
\usepackage{natbib}
\usepackage{multirow} 
\usepackage{float}
\bibpunct{(}{)}{;}{a}{}{,} % to follow the A&A style

\newcommand{\micron}{\mbox{$\mu$m}}

\begin{document}

	\title{The feedback of atomic jets from embedded protostars in NGC~1333}
	
	\author{Odysseas~Dionatos%\inst{1}
          \and
	   Manuel~G{\"u}del%\inst{1}
          }

\institute{
		University of Vienna, Department of Astrophysics,  T{\"u}rkenschanzstrasse 17,
A-1180, Vienna, Austria\\
		\email{odysseas.dionatos@univie.ac.at} \\
             }

%\titilerunning
\abstract
{[Abridged] Star-formation feedback onto the parent cloud is conventionally examined through the study of molecular outflows. Little is however known on the effect that atomic ejecta, tracing fast shocks, can have on the global cloud properties. In this study we employ Herschel/PACS  [\ion{O}{i}]  and  [\ion{C}{ii}]  spectral line maps of the NGC~1333 star-forming region to assess the relative influence of atomic jets onto the star-formation process. Atomic line maps are compared against molecular outflow tracers and atomic ejecta are associated to individual driving sources.  We study the detailed morphology and velocity distribution of  [\ion{O}{i}]  line using channel and line-centroid maps and derive the momentum, energy, and mass flux for all the bipolar jets traced by  [\ion{O}{i}]  line emission.  We find that the line-centroid maps can trace velocity structures down to 5 km s$^{-1}$ which is a factor of $\sim$20 beyond the nominal velocity resolution reached by Herschel/PACS. These maps reveal an unprecedented degree of details that assist significantly in the association and characterization of jets and outflows. Comparisons of the dynamical and kinematical properties shows that  [\ion{O}{i}]  momentum accounts for only $\sim$1\% of the momentum carried by the large scale CO outflows but the energy released through the jets corresponds to 50 - 100\% of the energy released in outflows. The estimated ratios of the jet to the outflow momenta and energies are consistent with the results of two-component, nested jet/outflow simulations, where jets are associated to episodic accretion events. Under this scenario, the energy from atomic jets to the cloud is as important as the energy output from outflows in maintaining turbulence and dissipating the cloud gas.}

\keywords{Stars: formation - Stars: jets - ISM: jets and outflows - ISM : kinematics and dynamics - ISM: atoms  - ISM: individual objects: NGC 1333}

\maketitle

%%%%%%%%%%%%%%%%%%%%%%%%%%%%%%%%%%%%%%%%%%%%%%%%%%%%%%%%%%%%%%%%%%%%%%

%\defcitealias{Plunckett:13a}{PACM13}

\section {Introduction}
\label{sec:1}

Protostellar jets and outflows are admittedly mechanisms holding keys in our understanding of star-formation for both isolated protostars but also on large, molecular cloud scales. Such ejecta are considered essential in removing the excess angular momentum from a protostellar system, allowing further accretion and therefore growth of the nascent star. Jets are well collimated structures, commonly associated with atomic ejecta from evolved protostars observed in the optical and near-infrared wavelengths, while outflows, typically associated with embedded sources, are traced in molecular lines forming wider, less collimated structures. While jets and outflows appear to co-exist to a greater or lesser extent in all phases of star formation \citep[e.g.][]{Nisini:15a, Podio:12a}, it is not yet clear how the two phenomena are linked to each other, or what is their relative influence to the star-formation process over time. There is currently growing observational evidence that outflow emission does not predominantelly trace jet-entrained gas but rather represents \textit{bona-fide} ejecta from the protostar \citep[e.g.][]{Davis:02a, Arce:13a}. Such evidence is also supported by a growing volume of simulations focusing on the early evolution of protostars \citep[e.g.][]{Machida:14a}.

The role of protostellar ejecta is not limited to regulating star-formation but as dynamical phenomena they also shape and influence the environment in the direction along their propagation. On small scales around individual cores, the outward motion of ejecta shapes the surrounding envelope carving hollow cavities, dispersing material and efficiently reducing the star-forming mass-reservoir \citep{Arce:08a}. Protostellar ejecta on larger scales carry momentum and deposit energy on the molecular cloud, feeding turbulent motions and increasing the internal energy of the system. Given that the majority of stars form in clusters \citep{Lada:03a}, protostellar ejecta can disperse the cluster gas \citep{Arce:10a} or trigger star-formation \citep{Foster:96a}, therefore having a negative or positive effect on the star-formation efficiency.

Studies of the star-formation feedback onto their parent clouds  are most commonly conducted with observations of molecular lines tracing gas in outflows with typical velocities between 10 and 50~km~s$^{-1}$ \citep[e.g.][]{Arce:10a, Plunckett:13a}. Protostellar jets, proceeding at velocities in excess of 100~km~s$^{-1}$ are often observed to pierce through dense clouds and propagate for several parsecs \citep[e.g.][]{Eisloffel:97a}. However the energy and momentum deposit of jets on cloud-scales remains largely unknown, especially in the case jets represent a parallel mass-loss process along with the outflows. Here  we present observations of [\ion{O}{i}]  and [\ion{C}{ii}] in the low-mass star-forming region NGC~1333 in Perseus  \citep[$d = 235 \pm 18$ pc,][]{Hirota:08a}, aiming to estimate the relative importance of protostellar jets in regulating star-formation both on scales of individual sources but also for the star-forming region as a whole.  We present a detailed study of the [\ion{O}{i}] morphology based on the high-velocity line-wings but also on a detailed study of the gaussian line-centroids. The latter method produces maps of unprecedented detail, which are subsequently employed for a thorough morphological study of jets in NGC~1333. Line-centroid maps allow the direct comparison of the [\ion{O}{i}] line emission distribution with molecular line-tracers of jet-induced shocks and outflows. Aiming to constrain the role of the atomic jets, we derive kinematic and dynamical properties for the [\ion{O}{i}] emission and we compare to the properties derived by a detailed study of the CO outflows in the region \citep{Plunckett:13a}. This is to our knowledge the first effort towards a uniform comparison between atomic and molecular ejecta in a single star-forming region using homogeneous datasets and consistent methods. Based on the comparisons to the CO data, we discuss the relative importance of jets in removing momentum and releasing energy from the protostellar systems, but also their role in transferring momentum and depositing energy on the parent cloud.

 \begin{figure*}[!ht]
\centering\
\resizebox{\hsize}{!}{\includegraphics{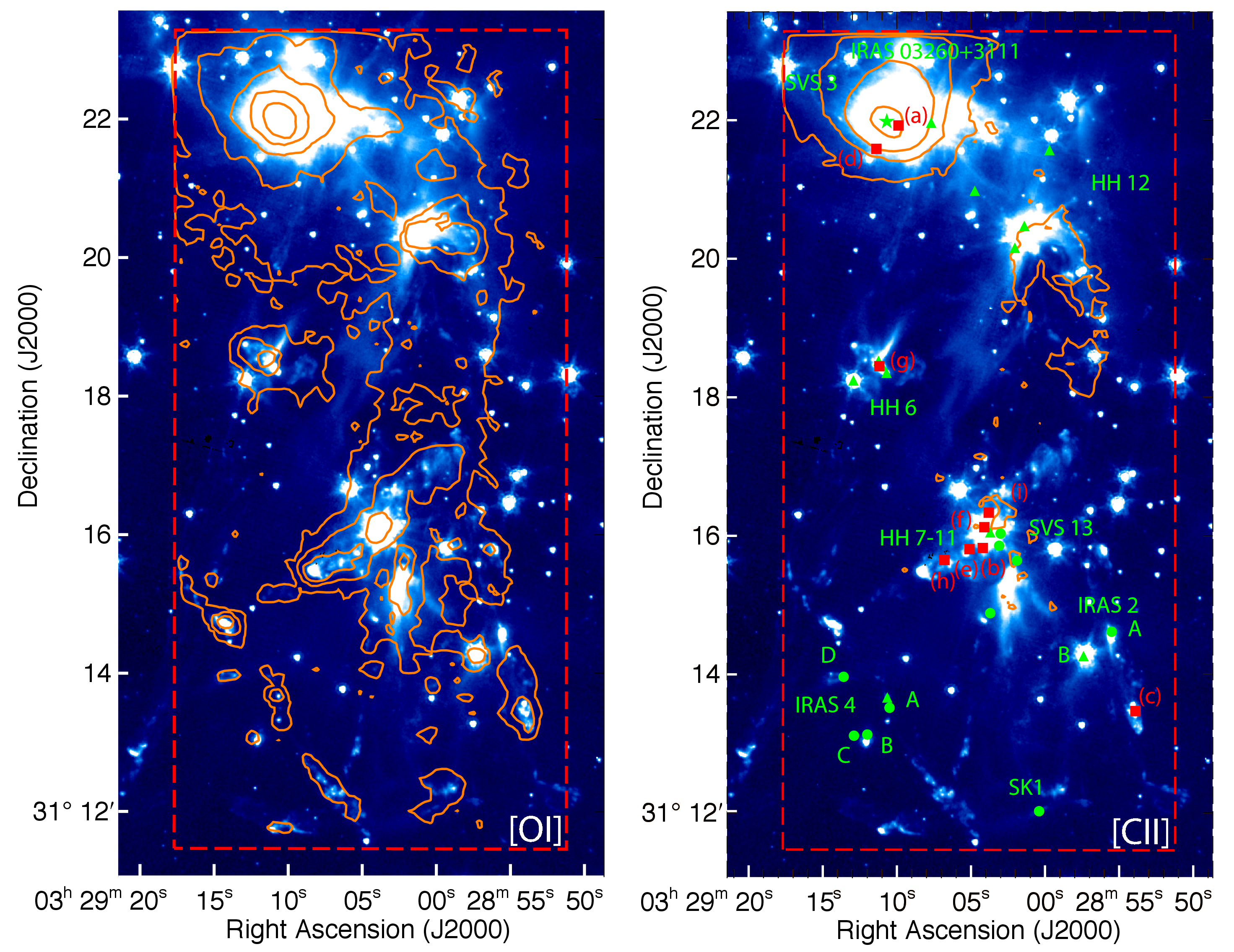}}
\caption{Contour maps of the integrated emission for the [\ion{O}{i}]  63 $\mu$m  and [\ion{C}{ii}] 158 $\mu$m lines (left and right panels, respectively), superimposed on an Spitzer/IRAC image centered at 4.5 $\mu$m. The (red) dashed line delineates the boundaries of the region mapped with Herschel/PACS. On the right panel the positions of the most prominent Class 0 (green filled circles), Class I sources (green filled triangles) and outflows discussed in the text are potted for reference. Filled (red) squares labelled (a - i) correspond to positions of spectra displayed in Fig.~\ref{fig:2}.  Contour levels start at 8 $\times$ 10$^{-15}$ erg cm$^{-2}$  s$^{-1}$ and increase logarithmically at steps of 0.5 dex.}
\label{fig:1}
\end{figure*}

\section{Data Reduction}\label{sec:2}

Observations of NGC~1333 were performed as part of the OT1 program entitled "A Herschel Study of Star Formation Feedback on Cloud Scales"  (H. Arce, P.I.) and data were retrieved from the Herschel Science Archive (HSA)\footnote{http://www.cosmos.esa.int/web/herschel/science-archive}. They consist of two maps obtained with the PACS spectrograph \citep{Poglitsch:10a} in line-scan  mode, centered at the rest wavelength of the [\ion{O}{i}] $^3$P$_2$ - $^3$P$_1$  and the [\ion{C}{ii}] $^2$P$_{3/2}$ - $^2$P$_{1/2}$ lines (63.185 and 157.741~$\micron$, respectively). Total duration of observations for each map is 30744.0 seconds ($\sim$ 8.5 hrs).  The [\ion{O}{i}]  and [\ion{C}{ii}] lines were observed with the B3A and R1-long PACS bands, respectively, in unchopped, line-spectroscopy mapping mode. [\ion{O}{i}]  and [\ion{C}{ii}] maps are centered at $\alpha_{J2000}$=03$^h$29$^m$03$^s$.3,  $\delta_{J2000}$=+31$^d$20$^m$49$^s$.9 and $\alpha_{J2000}$=03$^h$29$^m$03$^s$.2,  $\delta_{J2000}$=+31$^d$20$^m$39$^s$.8, respectively. Each map consists of tiling the 5~$\times$ 5 PACS footprint within a grid consisting of 8 raster lines and 17 raster columns, with a raster step of 41~$\arcsec$ in each direction. The total mapped area corresponds to $\sim$ 11.5$\arcmin \times$ 6$\arcmin$ and covers almost the entire NGC 1333 star-forming region.

Observations were processed with version 13 of the Herschel Interactive Processing Environment (HIPE), using the CalTree 69 pipeline, provided by the Herschel Science Center. At the end of the reduction/calibration process, individually mastered observations were combined to mosaics using the task \textit{specInterpolate}. The resulting mosaics have a spaxel size of 4.7$\arcsec$ with fluxes being interpolated using a Delaunay triangulation from the original input footprints, which are then projected onto an equidistant wavelength grid. 

After this point data were treated with custom scripts. Line  fluxes were calculated with simple integration on each spaxel after removing a first order polynomial baseline. In the case of the  [\ion{O}{i}] , line centroids were determined by Gaussian-fitting the emission-line components on each spaxel. Line-maps were reconstructed by re-projecting the calculated properties on a regular grid with a pixel size equal to that of the original cubes.   

The design values for the PACS spectral resolution at 63 and 157 $\micron$ correspond to R $\sim$ 3500 and 1200, which translate to a velocity resolution of $\sim$ 85 and 250 km s$^{-1}$ for the [\ion{O}{i}]  and [\ion{C}{ii}] line maps, respectively. In-flight calibrations and other observations have shown that in practice the spectral resolution values can be as low as 130 km s$^{-1}$ for the [\ion{O}{i}] line \citep[e.g.][]{Nisini:15a}. In this work we adapt an nominal resolution value of 100 km s$^{-1}$ for the [\ion{O}{i}] line. It should be noted that the \textit{specInterpolate} task oversamples the wavelength grid into 3.8~km~s$^{-1}$ wide bins, however, features at this velocity resolution shall be considered as artifacts.

\begin{figure*}[!ht]
\centering\
\resizebox{\hsize}{!}{\includegraphics{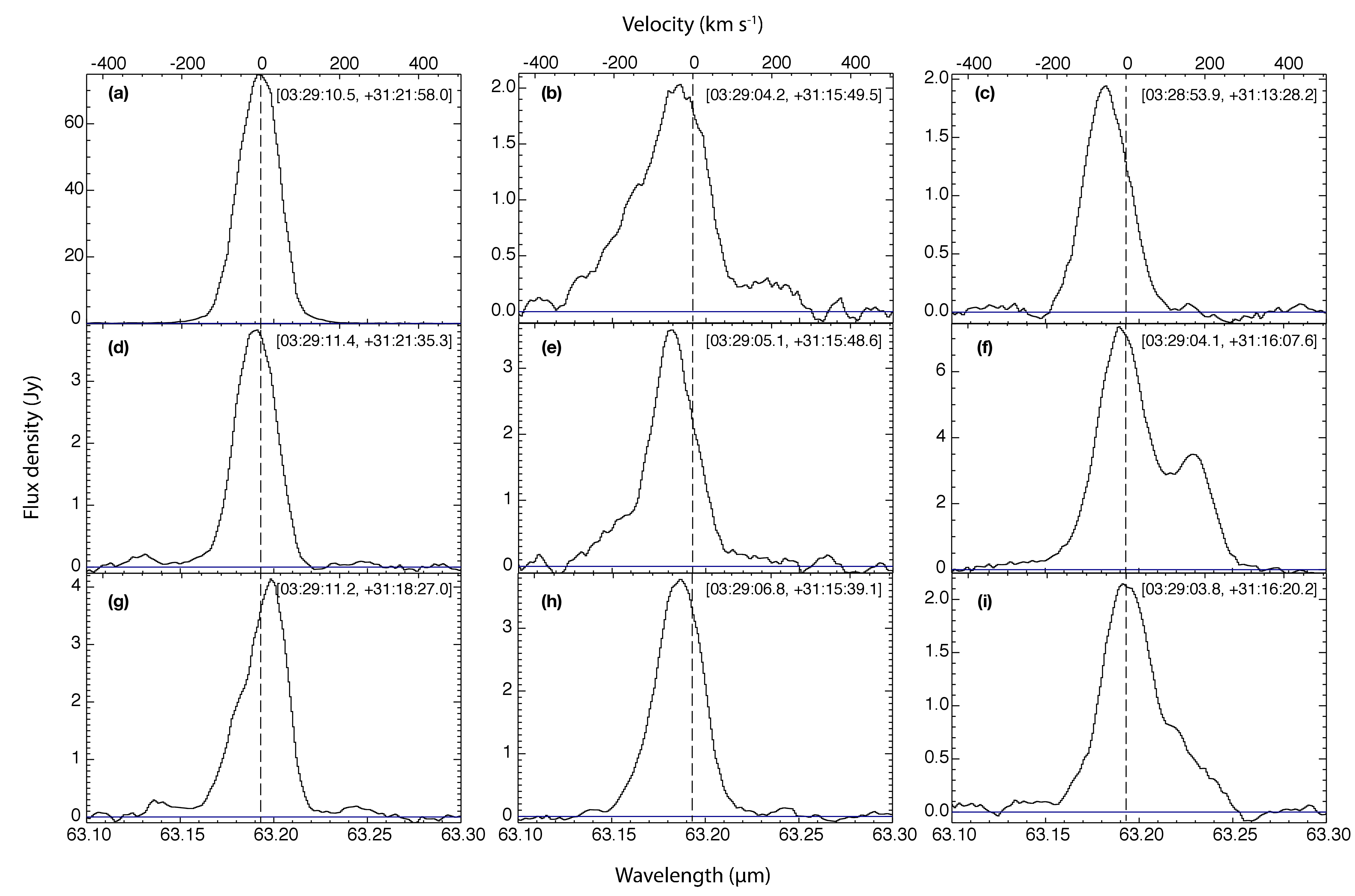}}
\caption{Continuum subtracted spectra of the [\ion{O}{i}]  line at 63 $\mu$m extracted from single PACS pixels on selected positions denoted with filled (red) squares on the right panel of Fig.~\ref{fig:1}. Coordinates corresponding to different locations are given on the top right of each panel. Dashed line indicates the [\ion{O}{i}] cloud velocity at 24~km~$s^{-1}$ or 63.191~$\micron$ (see also Fig.~\ref{fig:4}) }
\label{fig:2}
\end{figure*}

\section{Line emission}\label{sec:3}
 
 Here we discuss the spatial distribution of the [\ion{O}{i}]  and [\ion{C}{ii}] lines and through that associate their excitation to different processes. We examine the [\ion{O}{i}] line profiles and employ velocity-channel and line-centroid maps to study  different line-shape characteristics and their correspondence to the line maps.   
 
 \subsection{Overall morphology}

 The integrated line-emission maps of [\ion{O}{i}]  and [\ion{C}{ii}] are shown in Fig.~\ref{fig:1}, superimposed on a Spitzer IRAC band 2 image that is centered at 4.5~$\micron$. The emission in both maps is dominated by a strong peak towards the north-east, at a location coincident with the star IRAS~03260+3111(E). This  late  G-type star \citep{Doppmann:05a} is associated with the prominent reflection nebula SVS~3. The [\ion{C}{ii}] line at 157~$\micron$ is linked to energetic (X-ray and ultraviolet) radiation fields in photon-dominated regions (PDR's) \citep[e.g.][]{Dionatos:13a, Green:13a}, which is also suggested by diffuse near-infrared H$_2$ emission \citep{Greene:10a}.  While some morphological features as the jet-like pedestal extending to the west from IRAS~03260+3111(E) appear to have a [\ion{C}{ii}] counterpart, the large scale [\ion{C}{ii}] distribution in Fig.~\ref{fig:1} shows little or no association with the extended H$_2$ emission traced by the Spitzer/IRAC background image. The [\ion{C}{ii}] emission is dominant around in the north and remains prominent in the vicinity of  HH12 complex to the north-west, however it gradually fades towards the south and becomes barely detectable, with the exception of very localized patches of emission. The morphology of the [\ion{O}{i}] distribution in contrast shows significant structures and remains prominent across most of the mapped area. Both theoretical and observational studies associate the [\ion{O}{i}] emission as a shock tracer under the influence of protostellar jets \citep[e.g.][]{Hollenbach:89a, Nisini:15a, Green:13a, Dionatos:13a}. In the maps presented in Fig.~\ref{fig:1},  [\ion{O}{i}]  delineates \textit{almost} every single outflow, as traced by H$_2$ emission in the Spitzer/IRAC image so that its excitation can be attributed to the influence of shocks. There are very few exceptions to this phenomenology, where H$_2$ emission has no apparent [\ion{O}{i}] counterpart, however this may be related to the local shock conditions or the sensitivity of the [\ion{O}{i}] maps. The association between the [\ion{O}{i}] emission morphology with the H$_2$, CO and other jet/outflow tracers is discussed extensively in Sect.~\ref{sec:3.2}.   

Additional evidence on the association of the [\ion{O}{i}] to shocks induced by protostellar jets comes from the emission-line morphology, presented in Fig.~\ref{fig:2}. In this figure, the [\ion{O}{i}]  line-shape has a noticeable diversity corresponding to different locations across NGC~1333. The simplest shape (panel (a)  of Fig.~\ref{fig:2} ) corresponds to the emission around IRAS~03260+3111(E). The line appears almost symmetric but slightly shifted to the [\ion{O}{i}] rest velocity (more on the rest-velocity definition in the following Sect.~\ref{sec:3.1}) and has noticeable wings at the base.  Lines presented in all other panels of Fig.~\ref{fig:2} correspond to outflow regions, and their amplitude is a factor $\sim$~30 lower compared to panel~(a).  In all cases line-center is shifted with reference to the [\ion{O}{i}] rest velocity as much as  $\pm$~50~km~s$^{-1}$. Similar or even higher velocity shifts are reported in the study of [\ion{O}{i}]  emission around embedded protostars \citet{Nisini:15a} for the sources IRAS~4A and BHR~71. The [\ion{O}{i}]  line-morphology is often complex displaying multiple peaks (e.g. Fig.~\ref{fig:2}, panel~(f)), indicating the action of multiple jet components within the same spaxel. Signatures of high velocity gas appear as extended line-wings (e.g. panels (b, i)) or as separated-peaks, partly detached from the main line body (e.g. panels (d), (g)). The latter morphology is typically associated to high velocity bullets in outflows \citep[e.g.][]{Kristensen:12a, Dionatos:10b}. The velocity of the gas suggested by the extended line-wings reaches in some cases $\pm$~300~km~s$^{-1}$ (e.g. panel~(b)). 

\begin{figure*}[!ht]
\centering\
\resizebox{\hsize - 40pt}{!}{\includegraphics{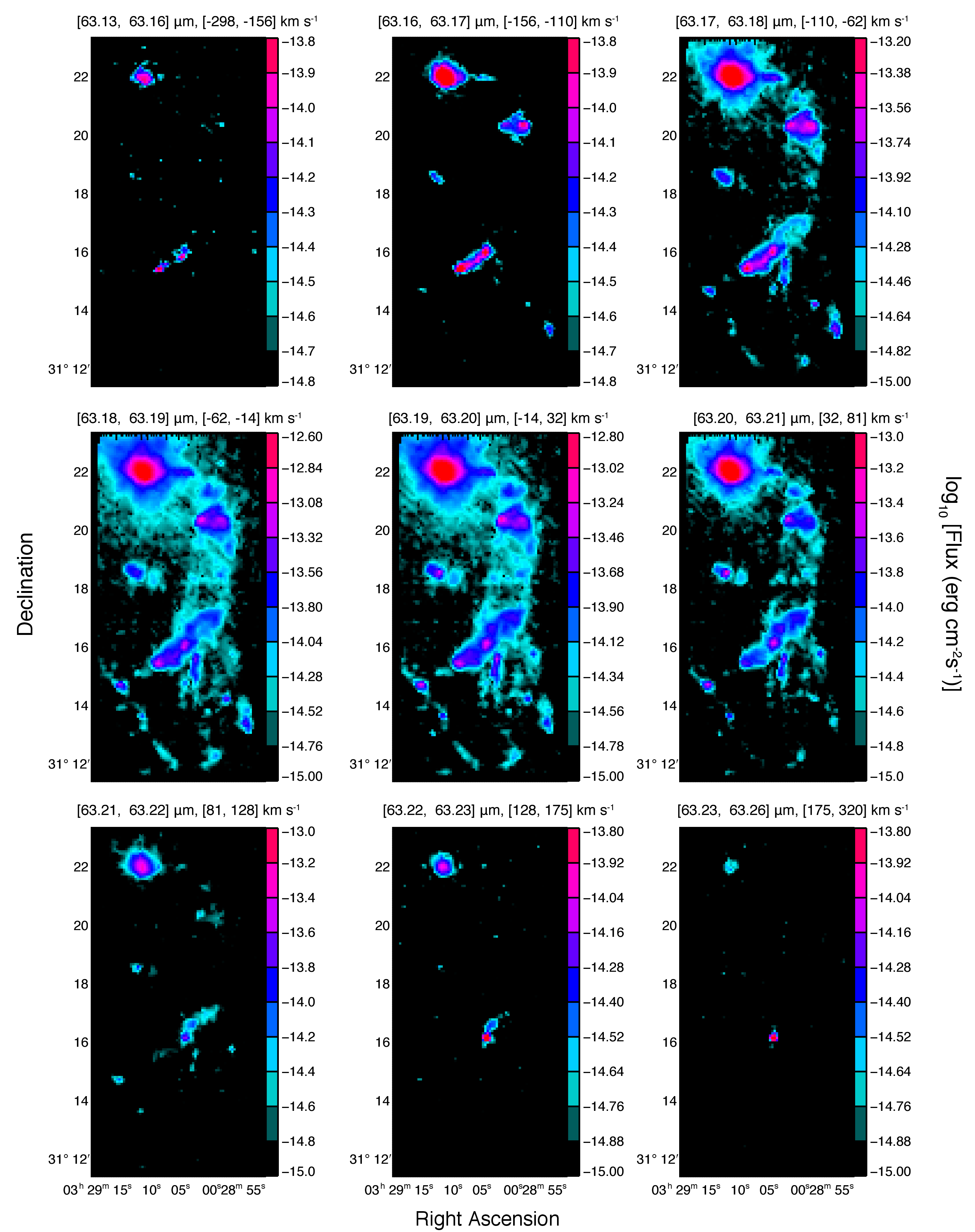}}
\caption{Velocity channel maps of the [\ion{O}{i}]  emission in NGC1333. For the extreme blue- and red-shifted velocity wings (top-left and bottom-right panels, respectively) line emission is integrated over 0.3 $\mu$m or $\sim$140~km~s$^{-1}$ while for the rest of the maps channel spacing corresponds to 0.1 $\mu$m or $\sim$ 45~km~s$^{-1}$. Integrated flux per channel is encoded in the corresponding color bars of each panel and levels are adjusted for maximizing the contrast and rendering weaker features visible.}
\label{fig:3}
\end{figure*}

Given the large velocity shifts observed across the map, we attempt to examine the spatial distribution of excited gas at different velocities in Fig.~\ref{fig:3}, notwithstanding the limited spectral resolution of the [\ion{O}{i}] line (see Sect.~\ref{sec:2}). The selected velocity bins are as a matter of fact coarse, ranging from 0.1~$\micron$ or $\sim$50~km~s$^{-1}$ around the body of the line and 0.3~$\micron$ ($\sim$150~km~s$^{-1}$) in the outer parts, in order to increase the signal-to-noise ratio in the high velocity wings. The strongest feature in the velocity channel maps (Fig.~\ref{fig:3}) remains the emission around SVS~3 which is prominent in all cases. For blue-shifted gas at highest velocities  [-298, -100]~km~s$^{-1}$ the most prominent features are associated with the HH~12 object to the north-west and the HH~7-11 outflow in the south. Symmetrical, red-shifted features can be identified between $\sim$ 80 and 300~km~s$^{-1}$ bins, which in the case of HH~7-11 stand out stronger. The morphology seen in the maps pertaining to velocities closer to the line-center is more confused.  Elongated blue- and red-shifted emission features corresponding to outflow lobes can be distinguished towards the IRAS~2 and IRAS~4 regions (to the south-west and south-east of the maps, respectively). The locations where the emission peaks  also show noticeable changes moving across different velocity-channel maps, indicating a significant shift in the location of the line-centers, where the peak of emission originates.

 \begin{figure}[!ht]
\centering\
\resizebox{\hsize}{!}{\includegraphics{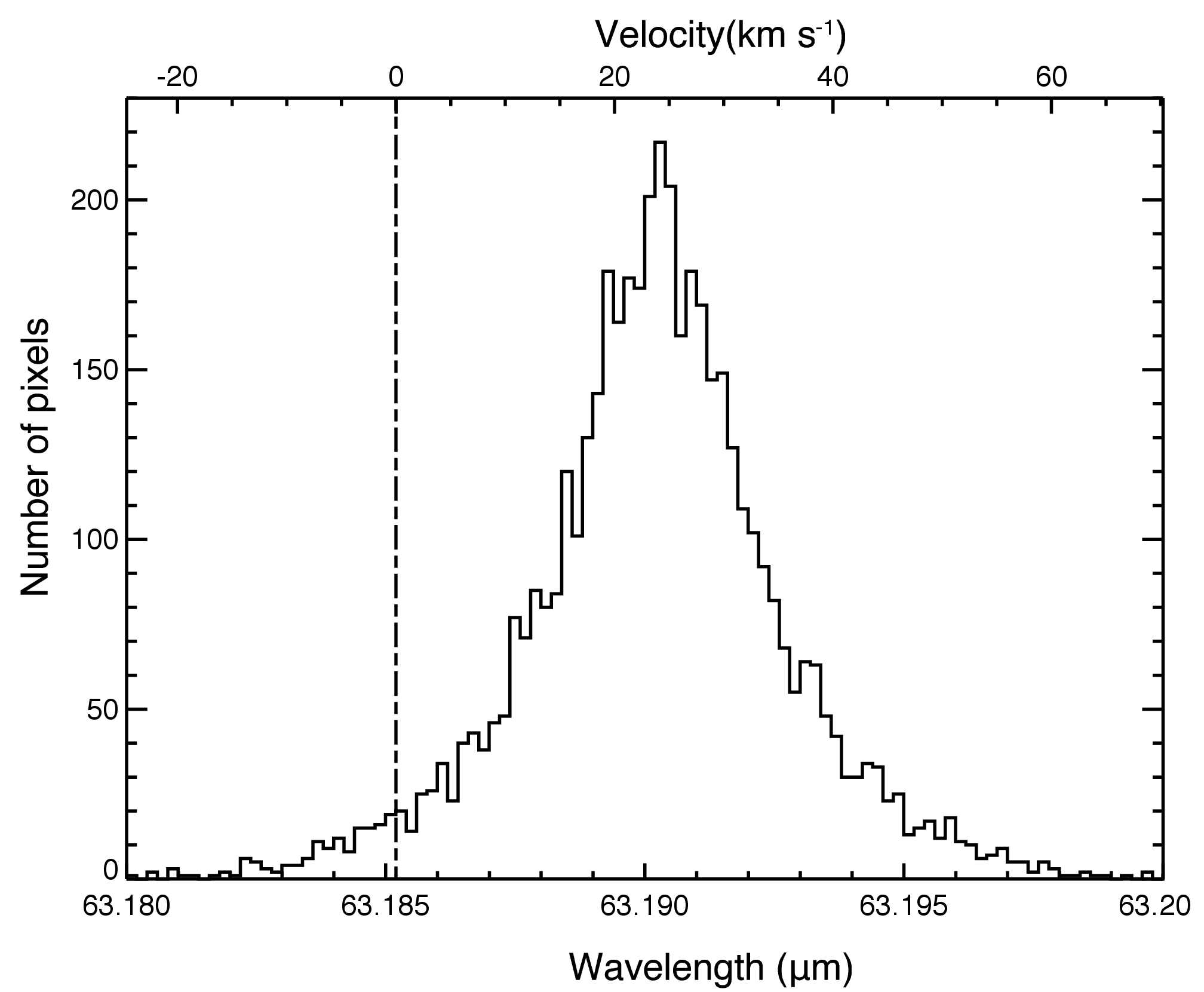}}
\caption{Number distribution of the [\ion{O}{i}]  line-centroids running for all spaxels of the NGC1333 map (see Fig.~\ref{fig:1}) in bins of 1~km~s$^{^-1}$. Dashed line indicates the location of the rest wavelength for the $^3$P$_2$ - $^3$P$_1$ transition of [\ion{O}{i}]  at 63.1852 $\micron$. The distribution is to a good approximation normal and peaks at 63,191 $\micron$, corresponding to a velocity shift of $\sim$~24~km~s$^{-1}$, which is taken as the [\ion{O}{i}]  rest velocity of the system.}
\label{fig:4}
\end{figure}
 
 \subsection{Gaussian line-centroids}\label{sec:3.1}

As discussed in the previous paragraphs, the [\ion{O}{i}] line-centers appear to be shifted to velocities of $\pm$50 km s$^{-1}$ or more (Fig.~\ref{fig:2}), a fact that also reflects in the spatial distribution of the [\ion{O}{i}] emission in the velocity channel maps of Fig.~\ref{fig:3}. We here extend the analysis of the velocity structures mapped with the  [\ion{O}{i}] line  by refining the determination of the position of the line-centers. To this end, each line is Gauss-fitted and the line-centroid is determined from the Gaussian peak centroid. The error in determination of the Gaussian peak line-centroids is given by the relation \citep{Garnir:87a}:

\begin{equation}
\sigma\approx  0.412 \sqrt{f/a}
\label{eqn:1a}
\end{equation}

\noindent
where $f$ is the full-width at half maximum (FWHM) of the line and $a$ is the relative line amplitude (the ratio of the line amplitude at the peak bin with respect to the full line). 
Taking as a useful resolution limit for our observations the instrumental FWHM of 100 km s$^{-1}$ (see also Sect.~\ref{sec:2}) and assuming a relative line amplitude of $a$ = 0.5, according to Eq.~\ref{eqn:1a} the error in the derivation of Gaussian peaks is of the order of 5.5~km~s$^{-1}$. This limit can be improved and reach below 5~km~s$^{-1}$ in cases the relative line amplitude is higher, as for example in the region around SVS~3 (see Fig.~\ref{fig:2}). We therefore find that line-centroids can retrieve details in velocity that are beyond the design capabilities of Herschel/PACS.  

For estimating the velocity centroid displacements, we first calculate the velocity shifts based on Gaussian-defined line-centroids for all spaxels of the [\ion{O}{i}] map and in Fig.~\ref{fig:4} we present their number distribution. In the same figure, the rest frequency of the [\ion{O}{i}] line at 63.185~$\micron$ (dashed line) also shown. The peak of the [\ion{O}{i}] line-centroid distribution lies at 63.192~$\micron$, which corresponds to a shift of 0.005~$\micron$ or a doppler-shift of $\sim$~24~km~s$^{-1}$. Examining the line-centroid distribution for different regions of the  [\ion{O}{i}]  map, we find that the peak of the distribution is dominated by the PDR-excited emission towards the north (above the dec= 31$\degr$:19$\arcmin$:00$\arcsec$ line) where emission lines are much stronger and therefore the line-centroids can be more accurately defined. Below that line, in the region that the emission is dominated by outflows, the distribution is less pronounced but remains symmetrical around the same velocity. 

The systemic velocity of NGC1333, as determined through a number of studies on water masers and outflows, is found to be +8~km~s$^{-1}$ \citep[e.g.][]{Rodriguez:02a, Choi:05a, Persson:12a}. This is a factor of 3 lower compared to the rest velocity of the [\ion{O}{i}] as defined through the number distribution of the line-centroids. The reason for this discrepancy is not yet clear. It is possible that systematics may have been introduced during the last phase of the data reduction, were the spaxels are projected onto an equidistant wavelength grid.  All steps of the reduction involving wavelength calibration of the PACS data were double-checked and to the best of our knowledge no problems were found. The [\ion{O}{i}] rest-velocity measurement, is in any case exact with respect to the observations discussed here, and is therefore used throughout the paper (see also the following Sect.~\ref{sec:3.2}).

\begin{figure*}[!ht]
\centering\
\resizebox{\hsize}{!}{\includegraphics{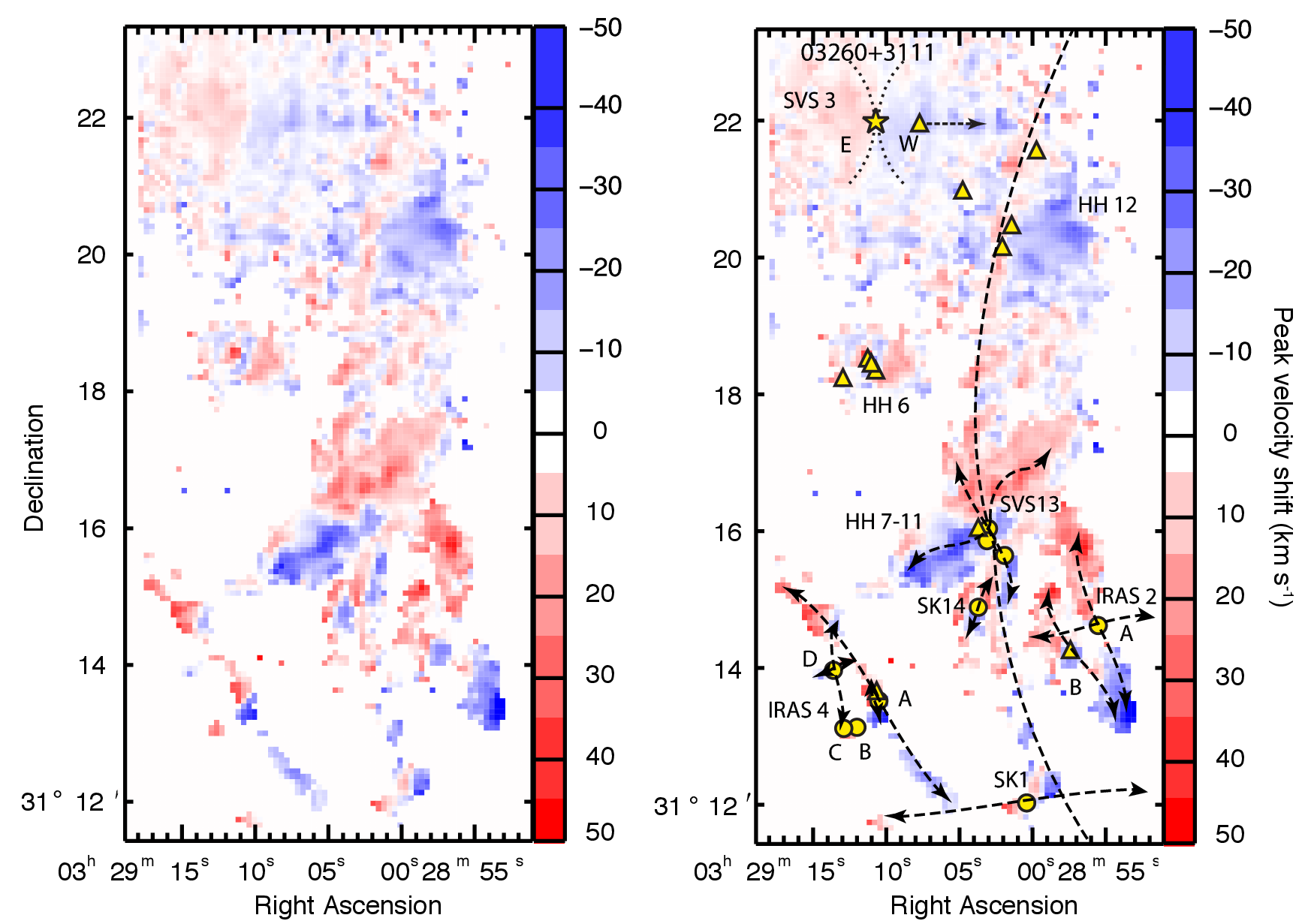}}
\caption{(\textit{Left:}) Map of the velocity shift observed in the [\ion{O}{i}]  line-centroids with respect to its rest velocity. Color encoding for each pixel corresponds to blue - and red- shifts, as indicated on the corresponding bar on the right side of the plot. (\textit{Right:}) A finder chart based on the left-hand side map, showing outflows (dashed arrows) and embedded (Class 0 and Class I ) sources (open circles and triangles, respectively). Dotted-X in the north shows the borderline between the low velocity blue- and red-shifted gas tracing the photodissociative wind from IRAS~03260+3111(E) (position of source marked with a star).  }
\label{fig:5}
\end{figure*}

\subsection{Line-centroid maps}\label{sec:3.2}

The spatial distribution of the [\ion{O}{i}] line centroids is presented in the left-panel of Fig.~\ref{fig:5}. To assist the discussion of the great amount of details revealed in this map we provide a finder-chart in the right panel of Fig.~\ref{fig:5}. Starting from the north at the top of the map and around SVS~3, the line-centroids reveal a slow $\pm$~10~km~s$^{-1}$ wind associated with the PDR region associated with IRAS~03260+3111(E). The emission appears to be red- and blue-shifted towards to the east and west direction with respect to the position of the system IRAS~03260+3111(E) with an axis of symmetry running almost in the north-south direction (denoted as dotted-X in Fig.~\ref{fig:5}). The velocities traced in the map are compatible with the velocities predicted to be generated by photo-evaporative winds in protostars \citep[e.g.][]{Woitke:09a, Gorti:09a}. This interpretation is also consistent with the maximum of [\ion{C}{ii}] line emission which is observed in the same region \citep[e.g.]{Greene:10a}. 

At the same declination and extending directly to the west, a series of  blue-shifted knots from the Class I protostar IRAS~03260+3111(W)  \citep{Evans:09a} delineates the prominent jet that appears also in the blue-shifted channels of Fig.~\ref{fig:3} (dashed arrow).  The morphology becomes more confused to the north-west around the region of HH12, where the bulk of the emission is blue-shifted. The cluster of high velocity (> 30 km s$^{-1}$) knots associated with the HH12 object correspond to a series of resolved knots observed in the near-IR H$_2$ emission \citep{Hodapp:95a}, but also with blue-shifted CO emission \citep{Knee:00a, Curtis:10a}. HH12 shows also significant [\ion{C}{ii}] emission extending to the south where the Spitzer/IRAC image depicts an emission-free cavity, which is likely associated with the Class I source SK26 \citep{Sandell:01a, Enoch:08a}. Class I sources are copious emitters of energetic photons \citep[e.g. X-rays,][]{Guedel:07a} which can excite and ionize the surrounding gas. The blue-shifted emission is interrupted only by a long arc-shaped stream of red-shifted knots running roughly in the north-south direction (dashed line), which is likely symmetric with a series of blue shifted knots extending to the far-south. This south-extending structure most likely corresponds to the extended, blue-shifted CO emission observed and attributed to SVS13B in \citet{Curtis:10a}, so in this configuration the structure represents a single, long outflow running in the north-south direction through whole [\ion{O}{i}] map.

Towards the center of the map and slightly to the east, the HH6 clump displays patchy, entangled blue- and red-shifted emission, linked to outflow activity from four Class I protostars embedded in this region. The angular resolution of the maps is not sufficient to disentangle individual outflows, however patches of shocked, H$_2$ emission towards the south have been kinematically associated to HH6 \citep{Raga:13a}.   

In south, located close to the map center, at the region where outflow emission dominates, the high velocity peaks in the velocity-centroid maps show an astonishing degree of symmetry between blue and red-shifted structures. The most prominent feature, HH~7-11, shows a series of high-velocity, blue-shifted knots that appear in reflection in the symmetric red-shifted lobe extending to the north-west. The area around the interface between the HH~7-11 and the red-shifted counter-lobe is packed with a number of Class 0 and Class I protostars driving overlapping and possibly interacting outflows (see also panel (f) in Fig.~\ref{fig:2}). The bipolar outflow as a whole is associated with the Class 0 source SVS13A, located at the center of symmetry between the lobes. The chain of high-velocity peaks follows a wavy, sinusoidal pattern, indicating a strongly precessing outflow. It can as well be that these structures correspond to a number of different outflows,  driven by more than one source. Other embedded protostars in the field are associated with prominent outflow structures. SVS13B is likely driving the large-scale outflow discussed in the previous paragraph, and SVS13C is associated with the bipolar outflow extending in the northeast-southwest (dashed arrows in Fig.~\ref{fig:5}, see also \citet{Plunckett:13a}). Directly to the south, we confirm the tentative association from \citet{Plunckett:13a} of a narrow bipolar outflow to the Class 0 source SK14. 

To the south-west, the line-centroids map reveals two, almost parallel bipolar outflow structures. The more extended of the two is associated with IRAS~2A outflow towards the north-south direction \citep{Plunckett:13a, Curtis:10a}. Attributing a driving source to the smaller bipolar structure is however not straightforward as there are no embedded sources  around the center of symmetry of this structure. The nearest embedded protostar is the Class I source IRAS~2B that is located at the same spot as the blue-shifted emission. We tentatively associate this outflow structure to IRAS~2B as presented in Fig.~\ref{fig:5}. In this configuration,  the projected directions of the blue-shifted emission from IRAS~2A and IRAS~2B overlap, noting however that different outflow configurations may be valid in the case IRAS~2B is an embedded binary protostar \citep{Tobin:16a}. A third outflow in the region that is associated with the IRAS~2A binary extending in the east-west direction \citep{Tobin:15a, Plunckett:13a} is not clearly detected in the line-centroid maps, with the exception of a red-shifted emission-patch at the tip of the red-arm. 

In the south-east, the narrow but extended jet-like emission is associated to the source IRAS~4A. The binary nature of the source \citep{Tobin:16a} may explain the different projected angle between the inner and outer high velocity knots.  The region around this corner of the map is speckled with a number of smaller, isolated, high-velocity clumps  which can be attributed in multiple configurations to a number of embedded sources lying in that region. Information on the proper motions as derived from the H$_2$ knots observed in the mid-infrared \citep{Raga:13a} is often complex, confusing, and in some cases suggestive that some of these clumps may be associated to larger outflow structures originating further up to the north, around HH~6 or further beyond. Based on pure symmetry assumptions, we tentatively associate four such clumps into two bipolar outflow schemes linked to IRAS~4D (see~Fig.~\ref{fig:5}).  Finally,  at the very south of the  [\ion{O}{i}] map and following the suggestion of \citet{Plunckett:13a} we associate a series of clumps in the east-west direction to the Class 0 source SK1.

\begin{figure*}
\centering\
\resizebox{\hsize}{!}{\includegraphics{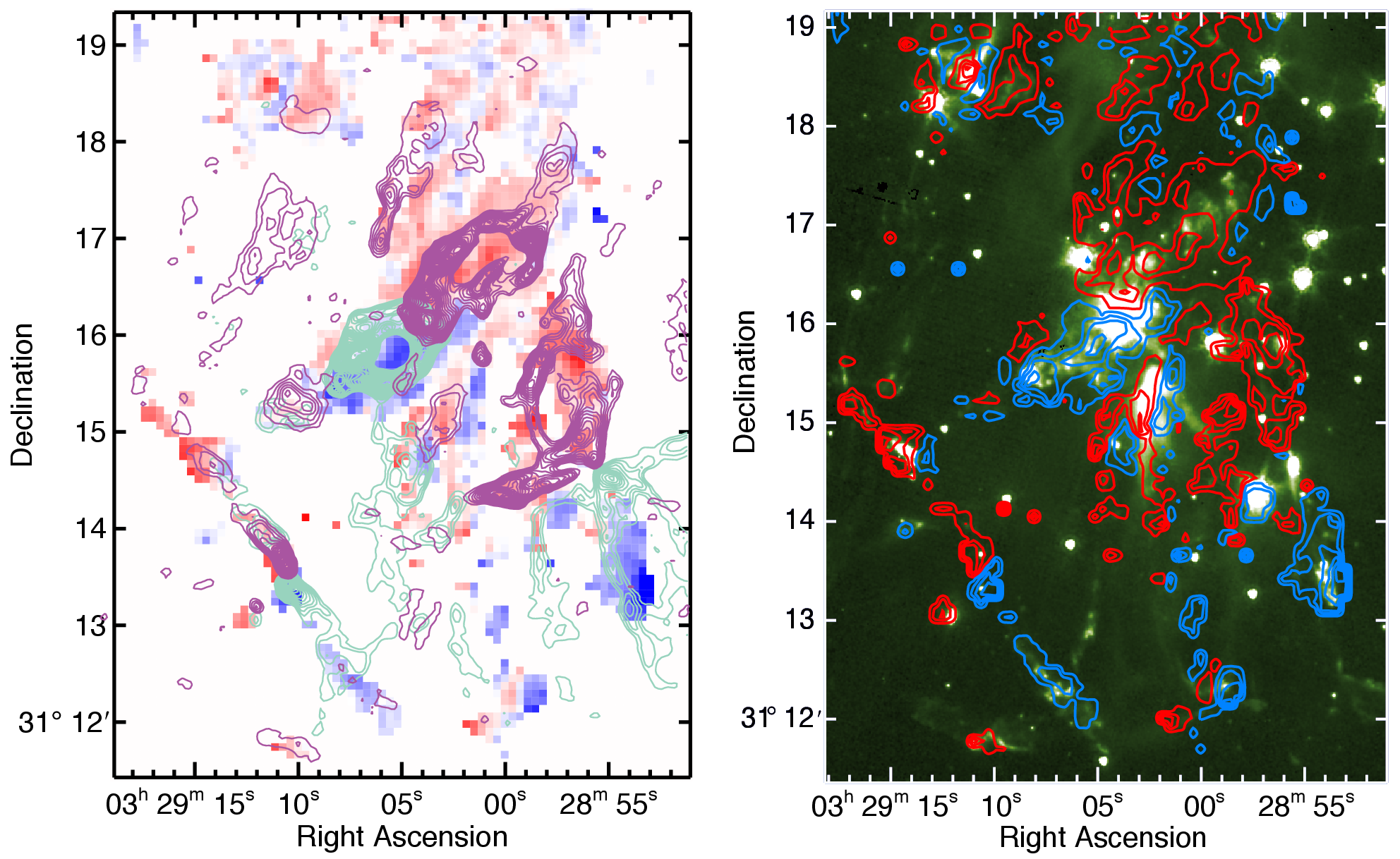}}
\caption{\textit{Left:}The outflow structures as recovered by the [\ion{O}{i}]  centroid analysis, in comparison with CO (1-0) interferometric map of \citet{Plunckett:13a}  Color levels are as  in Fig.~\ref{fig:5} and the frame focuses in the southern, star-forming complex of NGC~1333. \textit{Right:} [\ion{O}{i}] line-centroid velocity shift superimposed on an IRAC~4.5~$\micron$ image showing the close spatial coincidence between the two shock tracers. Contour levels start at 5 km s$^{-1}$ and increase by 10 km s$^{-1}$ increments for both blue- and red-shifted channels. }
\label{fig:6}
\end{figure*}

It is instructive to study the kinematics of the [\ion{O}{i}]  line-centroid maps in comparison with other common jet and outflow tracers in order to access the quality of the line-centroid method and role of the [\ion{O}{i}] emission as a jet/outflow tracer. In the left-hand panel of Fig.~\ref{fig:6}  we present the $^{12}$CO~$J =  1 \rightarrow 0$ outflow map of \citep{Plunckett:13a} obtained with CARMA, superimposed on the [\ion{O}{i}] velocity centroid map. The angular resolution of the CARMA map is $\sim$~5$\arcsec$,  therefore almost identical to the $\sim$~4.5$\arcsec$ spaxel scale of the [\ion{O}{i}] maps of this work. The velocity resolution of the CARMA maps, however, is a factor of 20 better than the limited reached with the velocity centroid method and can therefore trace reliably more detailed velocity structures. The maps shown in Fig.~\ref{fig:6} display very similar morphologies, tracing consistently every large scale outflow structure. The high velocity peaks [\ion{O}{i}] map fill-in the hollow, ``donut-shaped'' outflows lobes and cavities of the CO emission, rendering the two maps complimentary. An exception to this general trend is the red-shifted lobe extending to the north-east from IRAS~2A, which seems to correspond to an emission cavity as seen in both the H$_2$ and [\ion{O}{i}] emission. We note that the CO~$J =  1 \rightarrow 0$  observations of \citet{Plunckett:13a} pertain only low velocity gas (up to $\pm$~10~km~s$^{-1}$), while higher velocities have been reported for higher $J$ CO transitions \citep[e.g.][]{Bachiller:00a}. What comes as a surprise, is that the low velocity substructures in the [\ion{O}{i}] map, such as the entangled blue- and red-shifted emission at the tip of the HH~7-11 outflow or the similarly entangled emission to the NE lobe from IRAS~4A have an almost exact correspondence in the CO maps. In addition, CO clumps which are observed around the HH~7-11 outflow and denoted as F1 - F5 in \citet{Plunckett:13a} correspond to regions where the [\ion{O}{i}] changes velocity sign from blue- to red-shifted and \textit{vice-versa}. 

As mentioned earlier on, the distribution of the [\ion{O}{i}], with very few exceptions, closely follows the distribution of the mid-infrared H$_2$ emission (Fig.~\ref{fig:1}). In the right-hand panel of Fig.~\ref{fig:6} we present the line-centroid velocity distribution of the [\ion{O}{i}] in comparison to the H$_2$ emission. The [\ion{O}{i}]  emission is not only in agreement with the H$_2$ distribution but the [\ion{O}{i}]  velocity peaks are centered with great accuracy at the locations the H$_2$ knots. We note that the IRAS~2A east-west outflow which is one of the dominant bipolar outflows in CO not traced in [\ion{O}{i}]  has also no visible counterpart in the Spitzer H$_2$ image. It appears that the east-west flow from IRAS~2A produces no shocks and remains purely molecular to large distances, which is possibly associated to the very low momentum carried by this structure \citep[][see also Table~\ref{tab:1}]{Plunckett:13a}.  As a comparison, the very young outflow from HH~211 appears to be purely molecular at the base close to the protostar, but shows strong H$_2$ emission from shocks further out \citep[e.g.][]{Dionatos:10a, Tappe:08a}.   

Summarizing, the [\ion{O}{i}] emission appears to be complimentary, filling the CO emission in a jet-entrainment fashion, and follows closely the H$_2$ knot distribution, showing that it is excited in shocks. It is likely that [\ion{O}{i}] is produced in situ, breaking apart O-bearing molecules such as CO or H$_2$O in dissociative $J$-shocks \citep{Flower:10a}, and the correlation between the high-velocity [\ion{O}{i}] emission with the H$_2$ knots are in support of such an assumption. The same argument may explain the lack [\ion{O}{i}] emission in fainter H$_2$ knots observed in a few cases, indicating that either the shock velocity and density is not high enough or the shock may not be dissociative but continuous ($C$-type).  The  [\ion{O}{i}] distribution strongly suggests that almost every single shock observed in NGC1333 is either a pure $J$-type shock, or that there is at least a dissociative component at each shock location, which as a fact is consistent with a bow-shock topology \citep{Hollenbach:97a}. In this respect, outflows can represent entrained ambient gas, generated at a mixing layer interface along the sideways extending bow-shock flanks in an ``internal working-surface'' configuration \citep{Raga:93a}.

\begin{figure}
\centering\
\resizebox{\hsize}{!}{\includegraphics{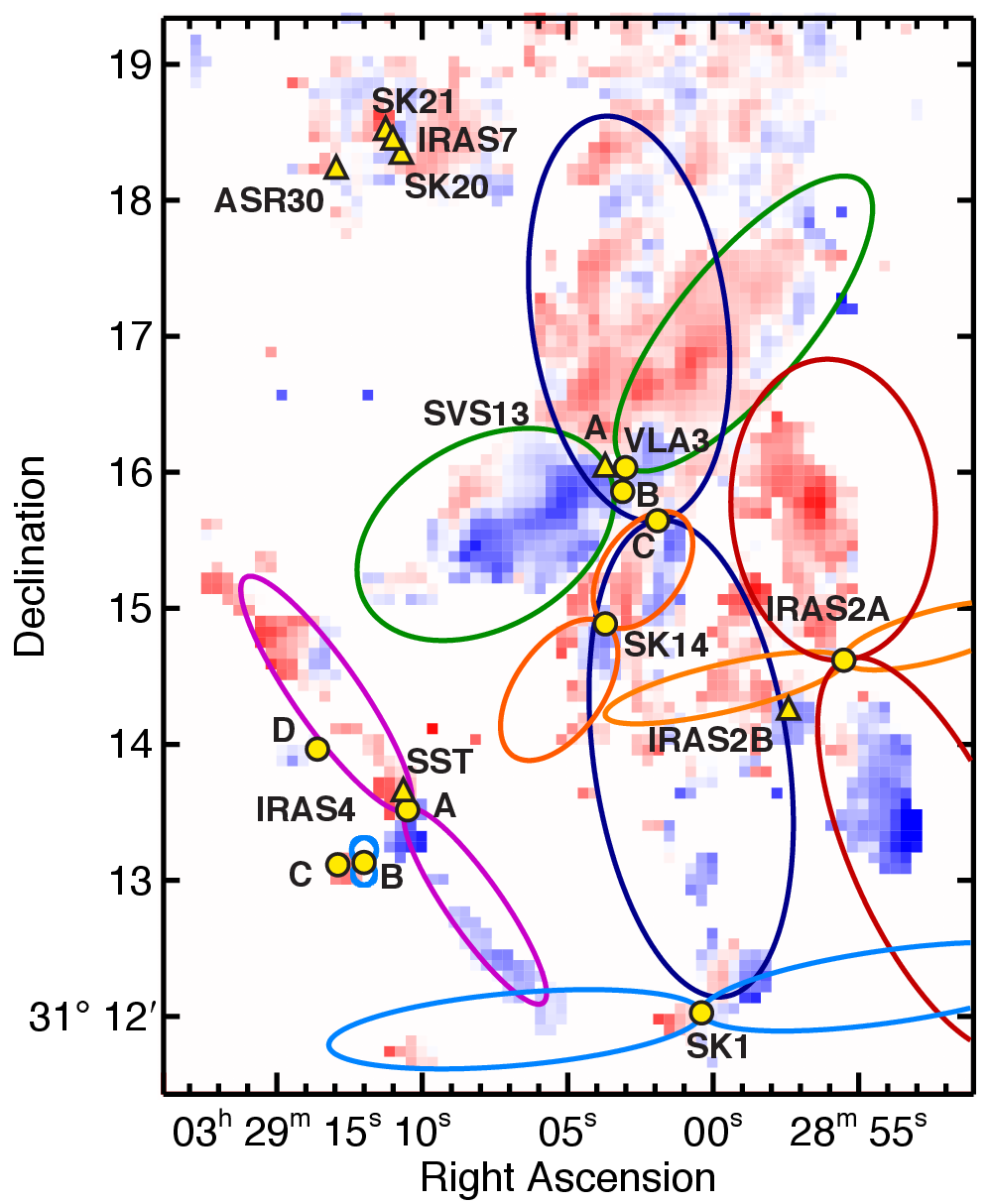}}
\caption{As in Fig.~\ref{fig:5} but focusing on the southern two-thirds of NGC~1333. Positions of embedded sources in the cloud are indicated with circles (Class 0) and triangles (Class I) and corresponding names are given for each source (the source labeled as \text{SST} next to IRAS4A corresponds to SSTc2dJ032910.65+311340.0). Colored ellipses delineate the outflow lobes attributed to embedded protostars in the work of \citet{Plunckett:13a}, and also employed here for the description of the outflow dimensions.}
\label{fig:7}
\end{figure}

\begin{table*}[!ht]
\small
\caption{Comparison of jet momentum and energies calculated from the [\ion{O}{i}]  line emission.}
\label{tab:1}
\centering
\begin{tabular}{l c c c c c c c c c   }

\hline\hline
Source  &  \multicolumn{5}{c}{Blue Lobe} &  \multicolumn{4}{c}{Red Lobe}  \\  \cline{2-5} \cline{7-10 } 
  & P$_{[\ion{O}{i}]}$ & P$_{CO}$$^{ a}$ & E$_{[\ion{O}{i}] }$ & E$_{CO}$$^{ a}$ & &  P$_{[\ion{O}{i}]}$  & P$_{CO}$$^{ a}$ & E$_{[\ion{O}{i}] }$ & E$_{CO}$$^{ a}$  \\
  & (10 $^{-2}$ M$_\odot$ km s$^{-1}$) & (M$_\odot$ km s$^{-1}$) & (10$^{43}$ erg) & (10$^{43}$ erg) &  & (10 $^{-2}$ M$_\odot$ km s$^{-1}$) & (M$_\odot$ km s$^{-1}$) & (10$^{43}$ erg) & (10$^{43}$ erg)  \\
%\hline
\hline

SVS13A & 4.5 & 2.7 & 10.0 & 9.8	 	& &		 3.7 &  2.2 & 7.7 & 7.8  \\
SVS13C & 1.8 & 2.5 & 4.2 & 9.7 		& & 		 5.0 &  3.3 & 10.1 & 12.1  \\
IRAS2A  W-E & 0.1 & 0.3 & 0.4 & 1.4 	& & 		  0.2  &  0.2 & 0.6 & 0.8  \\
IRAS2A  S-N & 0.8 & 1.8 & 2.1 & 8.1	 & & 	 0.9 &  1.3 & 2.7& 4.7  \\
IRAS4A & 0.3 & 0.3 & 0.6 & 2.3 		& & 		 0.8 &  0.4 & 1.5 & 1.4  \\
%IRAS4B & 10 & 0.1 & 0.1 & 0.02 		& & 		11 &  0.1 & 0.1 & 0.03  \\
SK14    & 0.2 & 0.3 & 0.4 & 1.3 			& & 		0.7 &  0.1 & 1.2 & 0.5 \\
SK1 & 0.2 & 0.2 & 0.5 & 2.5 			& & 		0.3 &  0.3 & 1.1 & 0.6  \\
\hline

\end{tabular}
\tablefoot{$^a $ CO outflow momenta and energies are from \citet{Plunckett:13a} and are reported here for comparison. }
%\footnote{a}{CO outflow momenta and energies are from the work of \citet{Plunckett:13a} and are reported here for comparison.}
\end{table*}

\section{The dynamics and kinematics of the [\ion{O}{i}] jets.  }\label{sec:4}

In the following we describe the methods for determining the mass, momentum and energy, but also the outflow dynamical timescales and the mass flux of the underlying protostellar jets, based on the  [\ion{O}{i}] emission maps. 

In order to compare with the CO $J = 1 \rightarrow 0$ emission  we calculate the [\ion{O}{i}] momentum for the outflows defined from the corresponding CO maps, using the reported source-position and outflow geometrical properties \citep[see Tables 2,3 and 4 in][]{Plunckett:13a}. We do not consider, however, any tentative outflows such as the flows C1 to C4 in the work of \citet{Plunckett:13a} and the monopolar flow from the protostellar source IRAS2B from the same paper. The geometry adopted from the CO outflows in \citet{Plunckett:13a}, superimposed on the [\ion{O}{i}] velocity map is shown in Fig.~\ref{fig:7}. The ellipses surrounding the CO outflows exceed in a couple of cases the limits of the [\ion{O}{i}] maps towards the west (Fig.~\ref{fig:7}), however this has negligible influence on our calculations as there is little  [\ion{O}{i}] emission extending in this direction.

\subsection{Mass, momentum and energy}\label{sec:4.1}

The momentum carried out by the [\ion{O}{i}] outflow is a key to our understanding of the impact of the atomic jets to the cloud environment.  Due to the large velocity dispersion observed in the [\ion{O}{i}] line (Fig.~\ref{fig:3}), we calculate the momentum using the following relation:

\begin{equation}
P_{[\ion{O}{i}] } = \sum_{v_{bin}} M_{v_{bin}}  \lvert v - v_{r} \rvert
\end{equation}
\noindent
where the sum acts over the mass per velocity bin, $v$ is the mean velocity for each velocity bin and $v_{r}$ is the rest velocity for the [\ion{O}{i}]  as defined in Sect.~\ref{sec:3.2}. The mass per velocity bin ($M_{v_{bin}}$) is derived from the  [\ion{O}{i}] luminosity per velocity bin, adopting the relation from \citet{Dionatos:09a}:

\begin{equation}
M_{v_{bin}} = \mu m_H \frac{L_{v_{bin}}[\ion{O}{i}]}{h\nu A_if_iX[O]}
\label{eqn:2}
\end{equation} 

In equation~\ref{eqn:2}, $\mu$ = 1.4 is the mean particle weight per H nucleus, m$_H$ is the mass of the hydrogen atom,  $A_i$ is the Einstein coefficient for spontaneous emission, $f_i$ is the relative occupation of level i, and X[O] is the abundance of atomic oxygen. The values for X[O] vary from 10$^{-3.52}$ to 10$^{-3.24}$ \citep[see][and references therein]{Rab:16a} and therefore introduce minor uncertainties in the mass calculations. A major contribution to the uncertainties involved in the mass derivation originate from the relative level occupation factor $f_i$ in the denominator of eq.~\ref{eqn:2}, which depends on the [\ion{O}{i}] excitation conditions, and in particular from the kinetic temperature, the density and the nature of the colliders. 

Constraining the excitation conditions requires observations of more than a single transition of an atom, and the only additional data available in this work comes from the [\ion{C}{ii}] maps. As discussed in Sect.~\ref{sec:3.1}, the 157~$\micron$ line is excited in radiative rather collisional processes and can therefore provide little information on the [\ion{O}{i}] excitation. For the following analysis we therefore need to rely on certain assumptions and employ results of similar studies found in the literature \citep[e.g.][]{Liseau:09a, Nisini:15a}. In a recent work employing [\ion{O}{i}] observations excitied in protostellar jets around five embedded sources with Herschel, \citet{Nisini:15a} employed non-thermal equilibrium (NLTE) models together with additional data on the $^3$P$_1$ - $^3$P$_0$ [\ion{O}{i}] line at 145~$\micron$ in order to study the possible excitation conditions of [\ion{O}{i}] . Summarizing the main results of this analysis, \citet{Nisini:15a} find that while the density of colliders is rather well constrained (ranging between $10^4 - 10^5$~cm$^{-3}$), the nature of the colliders (atomic or molecular hydrogen) can affect the 63$\micron$ [\ion{O}{i}] emissivity by a factor ranging between 2 and 5. In an order study, \citet{Liseau:09a} demonstrated that the emissivity of the 63~$\micron$ line can vary up to a factor of three for kinetic temperatures higher than 300~K, while in the same study they find that the excitation of the [\ion{O}{i}] 63~$\micron$ line typically traces kinetic temperatures of $\sim$~1000~K.

\begin{table*}[!ht]
\small
\caption{ [\ion{O}{i}] mass flux, jet dynamical timescales.}
\label{tab:2}
\centering
\begin{tabular}{l  c c c c c   c   c c c c c   }

\hline\hline
Source  &  \multicolumn{5}{c}{Blue Lobe} &  &\multicolumn{5}{c}{Red Lobe}  \\ \cline{2-6} \cline{8-12 } 
\\
  & $a^a $ & $v_t^b$& t$_{dyn}$ &$\dot{M}_{[\ion{O}{i}]}$ & $\dot{M}_{[\ion{O}{i}] - shock}$  &   & $a^a$ & $v_t^b$& t$_{dyn}$ &$\dot{M}_{[\ion{O}{i}]}$ & $\dot{M}_{[\ion{O}{i}] - shock}$   \\ \cline{5-6} \cline{11-12 } 
  & ($\arcsec$) & (km s$^{-1}$) & (10$^3$ yr)& \multicolumn{2}{c}{ (10$^{-7}$ M$_{\odot}/yr$)} &   & ($\arcsec$) & (km s$^{-1}$) & (10$^3$ yr)& \multicolumn{2}{c}{ (10$^{-7}$ M$_{\odot}/yr$)}  \\
%\hline
\hline

SVS13A & 122 & 20 & 6.8  & 1.1  & 19.8	 	& &		 163 &  20 & 9.1&0.8 & 18.3  \\
SVS13C & 212 & 100 & 2.4 & 1.4 & 9.3	 	& &		 180  &  10 & 20 &0.4 & 24.1  \\
IRAS2A W-E & 108 & 15 & 8.1& 0.01 & 0.01	 	& &    84 &  15 & 6.2 & 0.06 & 1.1  \\
IRAS2A S-N & 188 & 50 & 4.1  & 0.3 & 3.5	 	& &	  133 &  50 & 2.9 & 0.5 & 4.1  \\
IRAS4A & 104 & 100 & 1.2  & 0.4 & 1.1	 	& &		 125 &  100 & 1.4 & 1.0 & 3.8  \\
SK14 & 71 & 30  & 2.6 & 0.2 & 1.0	 	& &		 	58  &  30 & 2.2 & 0.7 & 4.2  \\
SK1 & 165 & 80 & 2.3 & 0.1 & 0.7	 	& &		     165 & 80 & 2.3 & 0.2 & 1.1  \\

\hline

\end{tabular}
\tablefoot{$^a $  Outflow lengths are adopted from \citep{Plunckett:13a}.\\
 $^{b} $Tangential velocities are adopted from \citet{Raga:13a}. }
%\footnote{a}{CO outflow momenta and energies are from the work of \citet{Plunckett:13a} and are reported here for comparison.}
\end{table*}

\begin{table*}[!ht]
\small
\caption{ [Properties of jets, outflows and their corresponding sources.}
\label{tab:3}
\centering
\begin{tabular}{l  c c c c c    c c c c c   }

\hline\hline
%Source  &  \multicolumn{5}{c}{Blue Lobe} &  &\multicolumn{5}{c}{Red Lobe}  \\ \cline{2-6} \cline{8-12 } 
Source & T$_{bol}^a$ & L$_{bol}^a$ & t$_{dyn}^b$ & P$_{[\ion{O}{i}]}^c$ & P$_{CO}^c$  & E$_{[\ion{O}{i}]}^c$  & E$_{CO}^c$ & $\dot{M}_{[\ion{O}{i}]}^c$ & $\dot{M}_{[\ion{O}{i}] - shock}^c$  \\
%  & $a$ & $v_t$& t$_{dyn}$ &$\dot{M}_{[\ion{O}{i}]}$ & $\dot{M}_{[\ion{O}{i}] - shock}$  &   & $a$ & $v_t$& t$_{dyn}$ &$\dot{M}_{[\ion{O}{i}]}$ & $\dot{M}_{[\ion{O}{i}] - shock}$   \\ \cline{5-6} \cline{11-12 } 
  & (K) & L$_{\odot}$ & (10$^3$ yr)& (10 $^{-2}$ M$_{\odot}$ km s$^{-1}$) &  (M$_{\odot}$ km s$^{-1}$)      & (10$^{43}$ erg)  & (10$^{43}$ erg)  & (10$^{-7}$ M$_{\odot}/yr$)& (10$^{-7}$ M$_{\odot}/yr$)  \\
%\hline
\hline

SVS13A & 250 & 59 & 8.0  &  8.2 & 4.6	 	& 	17.7	    & 17.6&1.9 & 38.1  \\
SVS13C & 36 & 4.9 & 11.2 & 6.8 & 5.0 	& 		14.3    & 21.8 &1.8 & 33.4  \\
IRAS2A W-E & 57 & 76 & 7.1 &  0.3 & 0.6	 	 &  1.0     & 2.1 & 0.07 & 1.2  \\
IRAS2A S-N & 57 & 76 & 3.5 & 1.7 & 3.7	 	 &	 4.8   & 12.8 & 0.8 & 7.6  \\
IRAS4A & 43 & 5.8 & 1.3 & 1.1 & 1.7	 	& 	2.1	     & 3.7 & 1.4 & 4.9  \\
SK14 & 59 & 0.2 & 2.4 & 0.9 & 0.7 	& 		1.6 	  & 1.8 & 0.9 & 5.3 \\
SK1 & 32 & 0.7 & 2.3 & 0.5 & 1.0	 	& 		1.6       & 3.1 & 0.3 & 1.8  \\

\hline

\end{tabular}
\tablefoot{$^a $ Source bolometric temperatures and luminosities along with the CO outflow momenta and energies are from \citet{Plunckett:13a} \\
 $^{b} $ Dynamical scale is the average from the two lobes, as reported in Table~\ref{tab:2} \\
 $^{c} $ All other outflow properties are the sum of their corresponding lobes.}
%\footnote{a}{CO outflow momenta and energies are from the work of \citet{Plunckett:13a} and are reported here for comparison.}
\end{table*}

As pointed out by \citet{Liseau:09a}, the 63~$\micron$ line is optically thin for H$_2$ column densities of $\sim$~$10^{19}$ cm$^{-2}$. On similar grounds, \citet{Nisini:15a} using the radiative transfer code RADEX \citep{vanderTak:07a}  estimate that an appreciable opacity change for the 63~$\micron$ line occurs for column densities larger than 10$^{22}$ cm$^{-2}$. Direct column density measurements for all major outflows in NGC~1333 using Spitzer spectral-line mapping of H$_2$ rotational transitions,  show a variation between  $\sim 10^{19}$ cm$^{-2}$ and $\sim 10^{18}$ cm$^{-2}$ for the ``warm'' and ``hot'' H$_2$ components, respectively \citep{Maret:09a}. Given the close association between the [\ion{O}{i}] and the H$_2$ emission (see Sect.~\ref{sec:3.2}), we can safely assume that there are no significant optical depth effects affecting the [\ion{O}{i}] 63~$\micron$ line so that the line emission probes the total volume of gas. Comparisons of the 63 and 145~$\micron$ [\ion{O}{i}] lines close to embedded protostars and further out show that there is no appreciable variation, which is interpreted as an indication that there is no significant extinction of the 63~$\micron$ line even deep within the protostellar envelopes \citep{Nisini:15a}. Summarizing, the mass derivation based on the [\ion{O}{i}] luminosity can vary up to a factor of $\sim$~10 due to the uncertainties in the local excitation conditions but there should be no appreciable losses due to extinction or optical depth effects.  

The jet energy traced by the [\ion{O}{i}]  line can be derived according to the following relation:

\begin{equation}
E_{[\ion{O}{i}] } = \frac{1}{2}\sum_{v_{bin}} M_{v_{bin}}  \lvert v - v_{r} \rvert^{2}
\end{equation}
\noindent
where the mass per velocity bin is defined in equation~\ref{eqn:2} and velocities follow the same notation as for the momentum. The uncertainties in the mass derivation discussed above also affect the estimation of the energy deposited by the outflows. The  derived [\ion{O}{i}] values for the momentum and energy are reported in in Table~\ref{tab:1} for each outflow source and lobe separately. In the same table and for facilitating direct comparisons we provide the corresponding CO momentum and energy values from \citet{Plunckett:13a}.

\subsection{Mass flux, dynamical timescale}\label{sec:4.2}

For the calculation of the mass flux, we follow the prescription of \citet{Dionatos:09a}, which has been also applied in a number of similar cases \citep[e.g.][]{Dionatos:14a, Nisini:15a}.  According to this, the mass flux can be estimated from the following relation:

\begin{equation}
\dot{M}_{[\ion{O}{i}]} =  \mu m_H \frac{L_{[\ion{O}{i}]}}{h\nu A_if_iX[O]}   \times t_{dyn}^{-1}
\end{equation}

Similar to the previous section, the mass determination is based on the [\ion{O}{i}] luminosity, with the difference that we now consider the total luminosity of gas confined within a given volume. The uncertainties related to the [\ion{O}{i}] excitation and the influence of optical depth  and extinction effects discussed in Sect.~\ref{sec:4.1} apply also in this case. For the estimation of the mass flux we define a dynamical scale from the projected length of the outflows and the tangential velocity of the emitting gas according to the following relation:

\begin{equation}
t_{dyn} = l_t/v_t
\end{equation}

Estimations of the tangential velocity rely on accurate proper-motion measurements that are often derived from different gas tracers than the ones employed for the derivation of the mass flux. In the present study we employ the measurements of \citet{Raga:13a} that are based on a multi-epoch study of H$_2$ knots observed with Spitzer. Given the close morphological correlation of the [\ion{O}{i}] and H$_2$ emission, we assume that the proper motions derived by the latter describe well also the [\ion{O}{i}] flow. Aiming to be consistent with the analysis of the previous paragraph, we adopt for the definition of projected outflow lengths the dimensions of the CO outflows from \citet{Plunckett:13a}.  The values of the parameters employed in the calculation of the dynamical timescales (projected length and tangential velocity), along with the derived values for the oxygen mass flux are reported in Table~\ref{tab:2}. In order to facilitate direct comparisons to the physical quantities of Table~\ref{tab:1}, values are reported separately for the blue- and red-shifted lobes.  
  
Alternatively, the mass flux of outflows can be directly estimated according to the relation showing that  the mass-loss rate is directly proportional to the [\ion{O}{i}]  luminosity \citep{Hollenbach:89a}:  
 
 \begin{equation}
  \frac{\dot{M}_{[\ion{O}{i}]_{shock}}}{ M_{\odot} yr^{-1}}= \frac{10^{-4} \times  L_{[\ion{O}{i}]} }{  L_{\odot}}
\label{eqn:6}
\end{equation}

This relation assumes  that $all$ [\ion{O}{i}] emission is generated in dissociative, $J-$type shocks,  an assumption that is observationally supported by an increasing volume of studies \citep[e.g.][]{Dionatos:13a, Nisini:15a}. As discussed in \citet{Nisini:15a}, the relationship considers a single shock interface where the velocity of the jet is assumed to equal to the shock velocity. Based on the velocities of jets from evolved protostars, \citet{Nisini:15a} assume that the shock velocity is in fact 3-6 times lower than that of the jet. This factor would then cancel the effect of multiple shocks within a jet lobe, so that  equation~\ref{eqn:6} holds as is and needs no further corrections. Values for the [\ion{O}{i}] mass flux estimated by the relation~\ref{eqn:6} are reported in Table~\ref{tab:2}, inline with previous estimations.

\section{Discussion}\label{sec:5}

The oxygen maps presented in the previous sections are of comparable quality in terms of angular and velocity resolution to the CO maps of \citet{Plunckett:13a}, which have been, to-date, of unprecedented detail, given the extent of the area covered. Within this area we consistently derived the dynamical and kinematical properties for the atomic and molecular components of seven bipolar configurations where jets and outflows can be strongly associated. For the analysis we followed constant definitions for the loci of molecular and atomic ejecta and used consistent methods on homogeneous datasets to derive their aforementioned properties.  Therefore, the analysis of this data provides us, for the first time, with a unique opportunity to study the relation between the atomic and molecular ejecta, but also to assess the influence of atomic jets in their immediate surroundings and estimate their feedback on large, star-forming core scales. 

\subsection{The relation between atomic jets and molecular outflows.}

\begin{figure*}[!ht]
\centering\
\resizebox{\hsize}{!}{\includegraphics{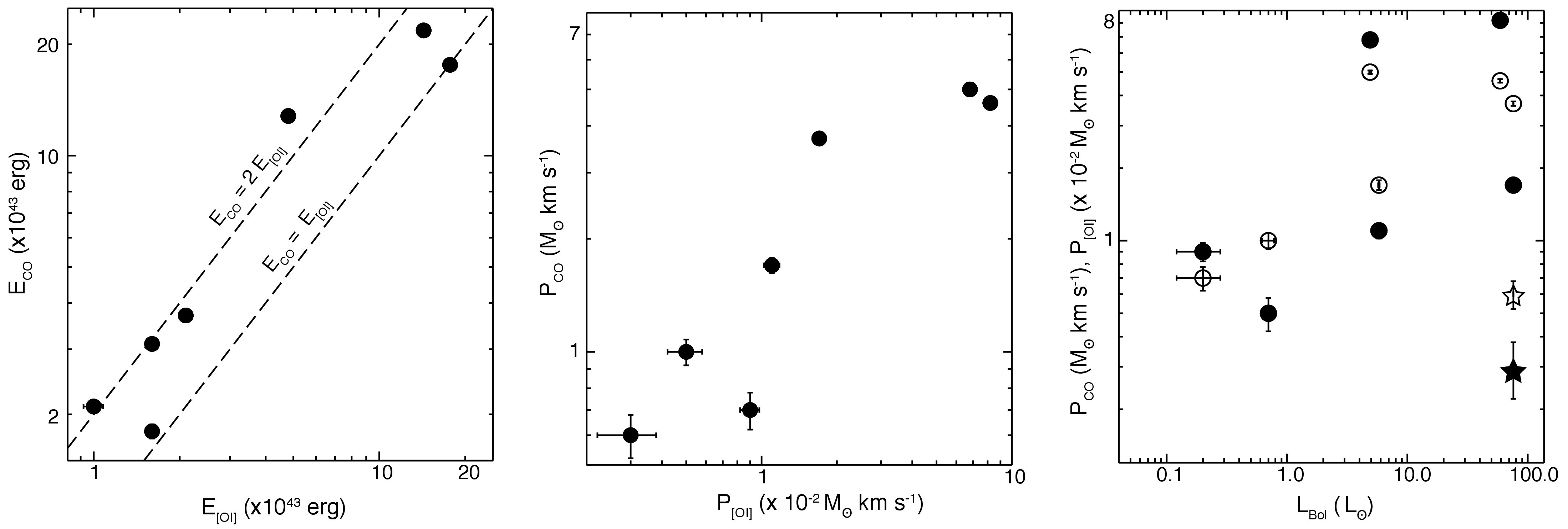}}
\caption{ Comparison of the energy and momentum (left and center panels, respectively) derived from the analysis of the [\ion{O}{i}] emission with the corresponding CO values from the work of \citet{Plunckett:13a}, as listed in Table~\ref{tab:3}. The CO and [\ion{O}{i}] momenta (empty and filled circles, respectively) are plotted against the embedded-source bolometric luminosities at the right-most panel. At the same panel, filled and empty stars correspond to the east-west outflow of source IRAS~2A. }
\label{fig:8}
\end{figure*}

The kinematic and dynamical properties of the bipolar [\ion{O}{i}] jets and CO outflows discussed in the previous sections along with the physical properties of their corresponding driving sources are summarized in Table~\ref{tab:3}. In Fig.~\ref{fig:8}, we compare the momentum and energy between the atomic and molecular ejecta (center and left panels, respectively). While the CO and  [\ion{O}{i}] momenta appear to closely correlate, the momentum carried along the  [\ion{O}{i}] represents only a fraction of $\sim$~1\% compared to the momentum corresponding to the CO emission. On the other hand, the energy carried by the atomic gas corresponds to 50-100~\% of the energy measured in molecular component. On the right hand panel of Fig.~\ref{fig:8}, we display how the momentum of the [\ion{O}{i}] and CO emission correlate with the bolometric luminosity of the driving sources.  We note that the star symbols correspond to the east-west directed outflow of the source IRAS~2A that is rather a peculiar case and is considered here as an outlier. While similar correlations for the momentum flux of molecular flows are well known \citep[e.g.][]{Bontemps:96a, Dionatos:10a}, this is the first time that such a correlation is shown to hold also for the [\ion{O}{i}]  ejecta emission related to embedded protostars and provides possible indications that the mass-loss phenomena also traced in atomic lines are directly linked to the accretion processes occurring very close to protostars. 

If we assume that the bona-fide ejecta from protostars along all evolutionary stages is atomic, as evidenced from the micro-jets of T-Tauri stars \citep[e.g.][]{Agra-Amboage:09a}, then the observed [\ion{O}{i}] emission can very well represent the same, atomic gas. In such a scenario, the molecular outflow emission traced in the CO lines would correspond to jet-entrained, ambient gas in the surrounding environment of forming stars.  While the physical structure and appearance of the molecular outflows would be dictated by the dominant modes of interaction between the underlying atomic jet with the surrounding medium \citep{Arce:07a}, a casual link between the two flows would translate in conserving the momentum. Given that the [\ion{O}{i}] velocity is $\sim$10 times higher than the CO gas, the [\ion{O}{i}] momentum is $\sim$~1\% of the CO and that the total energy is equally distributed between the two flows, then we find that the CO appears to be 10$^2$ - 10$^4$ times more massive than the [\ion{O}{i}] flow. In fact, the mass estimations in the previous paragraph are in support of the higher mass ratio. To a first approach, these results suggest that the atomic jet does not possess enough momentum to drive the observed CO outflows. Alternatively there may be significant [\ion{O}{i}] reservoirs in the atomic flow where the gas is not excited and therefore not detected, an assumption that can hold in the context of [\ion{O}{i}] excitation in shocks. In this scenario, the atomic gas is excited within a chain of shocks (internal working surfaces) and then cools radiativelly in the post-shock zone. Therefore the cumulative size of the regions where the emission is arising would be very small compared to the total length of the flow \citep[e.g.][]{Hollenbach:89a}. In this case, the  [\ion{O}{i}] emission measures only the gas excited in shocks but not the oblique atomic gas between shocks that is not radiating, but is still moving at relatively high velocities and therefore significantly contributes in stirring up a mixing turbulent layer that is observed in molecular line emission. In such a scenario the [\ion{O}{i}]  emission observed can indeed support the assumption that the protostellar ejecta are predominantly atomic.

In the discussion above we assumed that the atomic gas represents mainly ejecta from a protostar that is consequently excited in shocks. There is however enough evidence that the observed shocks are  dissociative, $J$-type, indicating that at least some of the atomic gas is produced \textit{in-situ} from the dissociation of oxygen-bearing molecules such as CO and H$_2 $O, at the location of the shocks. In such a scenario, the oxygen production would strongly depend on the shock-conditions such as the shock velocity and local density of the gas. Even though the shock conditions can have some affinity to the protostar, we would expect that the [\ion{O}{i}]  production and excitation in shocks would show little correlation to the properties of the forming star, which is in contrast with the correlation in the right panel of  Fig.~\ref{fig:8}. In that case the observed correlation would imply that each flow is directly linked to the driving source. The relation found between the momentum and the energy of the atomic and molecular ejecta is in agreement with the predictions of the numerical simulations performed by  \citet{Machida:14a} studying the formation and early evolution of protostars and taking into account two-component (jet and outflow) protostellar ejecta with a nested structure. In these simulations, the low-velocity outflow ($u_{out} < 50$ km s$^{-1}$) originates from the protostellar disk, is less collimated and is rather constant during the main accretion phase. On the other hand, the high-velocity jet is highly variable due to episodic accretion events which can be created by a number of different instabilities in the accretion disk (e.g. gravitational instability, \citet{Vorobyov:06a}, or a combination of gravitational instability and magnetic dissipation \citet{Machida:11a}). In this configuration, \citet{Machida:14a} demonstrates that the momentum carried out by the episodic jet outbursts is always much smaller than the more constant outflow, however the kinetic energy of the jet has spikes exceeding the energy of the outflows. As a result, the time-averaged values the jet kinetic energy can account for a significant fraction of the outflow energy, but in contrast the jet momentum, even at the highest peaks is too low to reach the outflow values.

\subsection{The feedback of atomic jets onto the star-forming core}

According to the properties measured from the [\ion{O}{i}]  emission,  the atomic jet momenta can have little influence on the cloud compared to the contribution form outflows. The energy deposited by the atomic jets, however, cannot be neglected. Summing up the jet and outflow energy output, the total energy feedback from protostellar ejecta to the cloud is $\sim 1.1 \times 10^{45}$ erg, or about 50\% higher compared to the total energy transported just by outflows \citep{Plunckett:13a}. This value compares directly to the turbulent energy of the cloud ($E_{turb} = 1.8 \times 10^{45}$ erg) estimated in \citet{Plunckett:13a} without accounting for the projected outflow inclination angles, and reveals the importance of the contribution of shocks in the cloud turbulence.  Following the analysis of \citet{Plunckett:13a}, the assumed dynamical timescale in the case of jet-driven outflows ($5 \times 10^4$ yr) is in excellent agreement with the mean of our estimates from Table~\ref{tab:3}. We find that the total luminosity from protostellar ejecta (outflow and jets/shocks) $L_{ej} = E_{ej}/t_{dyn} \approx 7 \times 10^{42} $ erg s$^{-1}$. The turbulent dissipation rate is given by $L_{turb} = E_{turb} / t_{diss}$, where $t_{diss} = 5.7 \times 10^5$ yr  is the energy dissipation time \citep{Arce:10a}. Comparing these two values we find a ratio $L_{ej}/L_{turb} \sim 7$ which is in support of the conclusions from \citet{Plunckett:13a} that outflows, but also jets, posses more than enough power to maintain turbulence, at least within the limits of the region studied here.

\subsection{The mass flux from the atomic jets}

The outflow mass flux derived Sect.~\ref{sec:4.2} is in good agreement with the values derived in \citet{Nisini:15a} for a number of outflows in different clouds. The outflow from IRAS~4A mass flux derived in \citep{Nisini:15a} is $\sim 1.5 \times 10^{-7}$ M$_{\odot}$~yr$^{-1}$, in excellent agreement with the values reported in Table~\ref{tab:3}. On average mass-flux values for NGC~1333 lie on the lower end compared to the values reported in \citet{Nisini:15a}.  This difference can be attributed on the one hand to the lower tangential velocities adopted from the work of \citet{Raga:13a}, and on the other hand to our assumption of fully atomic hydrogen as the main excitation agent. The mass flux estimations for the outflow from the source SVS13A,  based on the near-infrared [\ion{Fe}{ii}] and H$_2$ lines \citep{Davis:11a} range from $\sim 5 \times 10^{-7}$ to $10^{-6}$ M$_{\odot}$ yr$^{-1}$, comparable to the values we derive here with the two different methods. It should be noted however that the values reported \citep{Davis:11a}  are corrected for an average outflow inclination angle of 57$\degr$.3 and extinction. A general trend observed in the current analysis is that the the shock-derived mass flux can be up to a factor of 20 higher compared to the mass flux values calculated assuming a dynamical timescale. This is especially pronounced in cases of extended and confused lobes (e.g. in the cases of SVS13A and SVS13C), suggesting a higher-than-assumed number of shocks within each lobe, that is not effectively canceled by the difference between the jet and shock velocities, as assumed in \citet{Nisini:15a}.  

Assuming that the [\ion{O}{i}] emission represents bona-fide ejecta from the protostar, then we would expect to observe a clear trend in the mass flux of the atomic gas as a function of the evolutionary stage of a protostar, which would reflect the observed drop in accretion rates from Class 0 to Class II sources.  In a study of the [\ion{O}{i}] 63~$\micron$ emission with Herschel around five Class II protostars, \citet{Podio:12a} report mass fluxes in the between 10$^{-8}$ and 10$^{-6}$ M$_{\odot}$~yr$^{-1}$ with an average value of $\sim 10^{-7}$ M$_{\odot}$~yr$^{-1}$. Compared to the values reported here but also in \citet{Nisini:15a}, the jet mass flux shows if any, only a minor decline between the embedded and the T Tauri phase. These results are in support of  the two-component ejecta scenario of \citet{Machida:14a}, which is also supported  by the difference between the momentum and energy between the jet-shocked emission and the outflows as discussed in the previous sections.  

\section{Conclusions}\label{sec:6}

We presented Herschel/PACS line maps of the NGC~1333 star-forming region based on line-scan observations of the  [\ion{O}{i}] and  [\ion{C}{ii}] lines at 63~$\micron$ and 157~$\micron$, respectively. The emission from both tracers peaks towards the north of the mapped region due to the energetic radiation from the source IRAS~03260+3111(E), exciting the SVS~3 reflection nebula. Significant [\ion{C}{ii}] emission is also associated with the HH12 outflow region to the north-west. Towards the southern of NGC~1333, [\ion{C}{ii}] the emission becomes less important and the region is dominated by [\ion{O}{i}] emission excited under the influence of jets from young protostars. 

Our analysis focused in deriving the morphological, dynamical and kinematic properties of the atomic gas in the outflows as traced by [\ion{O}{i}]. We then compared these properties to the ones traced by other outflow tracers such as CO. Our main results can be summarized as follows:

\begin{itemize}

\item{Oxygen lines display a high morphological diversity reflecting gas moving at a wide range of velocities. Extended line wings trace in some cases velocities up to 300 km s$^{-1}$. The continuous or semidetached shape of the line-wings shows that the high velocity gas can be associated to smoothly accelerated structures or individual ``bullets'' of gas.}

\item{Notwithstanding the low velocity resolution of the [\ion{O}{i}] maps  velocity channel maps at bins of 50~km~s$^{-1}$  reveal outflow structures extending symmetrically to known protostellar sources.  Channel maps show also the significant shifts in the [\ion{O}{i}] line peaks, contributing to the local maxima in the different velocity bins.}

\item{A thorough estimation of the line-centroids as derived from Gaussian-fits to the [\ion{O}{i}] lines provided velocity structures down to 5 km s$^{-1}$, which is a factor of $\sim$20 higher compared to the design-capabilities of Herschel/PACS.}

\item{The maps based on line-centroids trace velocity structures down to 5~km~s$^{-1}$  exposing even slow,  symmetric flows from photo-evaporative winds. }

\item{The outflow structures  revealed in the line-centroid maps show a great degree of detail and symmetries, providing important information necessary to associate individual flows to driving sources. Based on these maps we confirm a large number known, and propose a few new bipolar outflow structures and their associated driving sources.}

\item{The resolving power of the line-centroid maps in small, low velocity structures is comparable to CO interferometric maps of much higher velocity resolution, as revealed by copious comparisons. At larger scales the oxygen and CO emissions reveal very similar but often also complimentary outflow structures.}

\item{The [\ion{O}{i}]  line centroid velocity peaks follow closely the H$_2$ emission indicating that oxygen is excited in shocks shocks. }

\item{The momentum carried by the [\ion{O}{i}] represents only the $\sim$~1\% of the momentum corresponding to the large-scale CO emission. }

\item{The energy traced by [\ion{O}{i}]  accounts for up to 100\% of the energy measured in CO outflows.}

\item{Assuming that the oxygen emission represents jets from protostars which are also responsible for driving the CO outflows, then there must be a significant amount of atomic jet mass that is not excited in shocks. This can be consistent if the space between a chain of shocks (internal working surfaces) is filled with high velocity atomic gas which is not observable but creates interacts with the surrounding medium flow instabilities, creating much of the observed CO emission.}

\item{The momentum and energy estimates are in also agreement with the predictions of a nested jet/outflow structure simulations of \citet{Machida:14a}, where the jet is linked to episodic accretion events, while the large-scale outflow is constant (non-episodic) and carries the bulk of the momentum.}

\item{The mass flux carried by the [\ion{O}{i}] jets, estimated for evolved Class II sources shows no significant decline when compared to the values corresponding to embedded protostars, estimated in this work. This finding may give additional support of the two-component ejecta scenario discussed above. }

\end{itemize}

\begin{acknowledgements}

This research was supported by the FFG ASAP-12 project \textit{JetPro*} (FFG-854025) and in part by the EU FP7-2011 project under Grant Agreement nr. 284405.  

 \end{acknowledgements}

\begin{tiny}

\bibliographystyle{aa}
\bibliography{ngc1333_pacs} 

\end{tiny}

\end{document}